\documentclass[aps,prl,showpacs,onecolumn,preprint,a4paper,superscriptaddress]{revtex4}

\usepackage{graphicx}
\usepackage{subfigure}
\usepackage{amsmath,amssymb}
\usepackage[nice,tight]{units}
\usepackage[latin1]{inputenc}
\usepackage[T1]{fontenc}
\usepackage{multirow}
\usepackage{color}

\newcommand {\de} {\mbox{d}}
\newcommand {\eref}[1] {(\ref{#1})}

\newcommand{\mytodo}[1]{}

\newcommand{\rem}[1] {}

\begin{document}

\begin{center}
 {\Large Soft Capacitors for Wave Energy Harvesting} \\
 \vspace{10ex}
 Karsten Ahnert,$^{1,2,*}$, Markus Abel,$^{1,2}$,\\
 Matthias Kollosche,$^{1}$, Per J{\o}rgen J{\o}rgensen,$^{3}$, and\\
 Guggi Kofod$^{1}$
\end{center}
\vspace{10ex}
$^{1}$ Department of Physics and Astronomy, Potsdam University, Karl-Liebknecht-Str. 24/25, D-14476, Potsdam-Golm, Germany \\
$^{2}$ Ambrosys GmbH, Geschwister-Scholl-Str. 63a, 14471 Potsdam, Germany \\
\vspace{2ex}$^{3}$ UP Transfer GmbH, Am Neuen Palais 10, 14469 Potsdam, Germany \\
\vspace{4ex}$^{*}$ Corresponding author, karsten.ahnert@ambrosys.de, Tel: +49 331 9775986\\
\noindent Major Subject: Physical sciences\\
\vspace{4ex}\noindent Minor Subject: Applied Physical Sciences \\
\noindent Date: \today

\newpage

\mytodo{Kommas in Aufzaehlungen pruefen: Das Komma vor and ist optional. (Entscheide dich für eine Variante, bleib im Text aber dabei.)}

\section*{Abstract}

Wave energy harvesting could be a substantial renewable energy source without impact on the global climate and ecology, yet practical attempts have struggled with problems of wear and catastrophic failure. An innovative technology for ocean wave energy harvesting was recently proposed, based on the use of soft capacitors. This study presents a realistic theoretical and numerical model for the quantitative characterization of this harvesting method. Parameter regions with optimal behavior are found, and novel material descriptors are determined which simplify analysis dramatically. The characteristics of currently available material are evaluated, and found to merit a very conservative estimate of 10 years for raw material cost recovery.

\newpage

\section{Introduction}

The problem of adequately supplying the world with clean, renewable energy is
among the most urgent today. It is crucial to evaluate alternatives to
conventional techniques. One possibility is energy harvesting from ocean waves,
which has been proposed as a means of
offsetting a large portion of the world's electrical energy demands \cite{Cruz-08}.
However, the practical implementation of wave energy harvesting has
met with obstacles, and the development of new methods is
necessary~\cite{Dalton2010}. Oceanic waves have large amplitude fluctuations that cause devices to fail due to excessive wear or during storms. A strategy to overcome these catastrophic events could be to base the harvesting mechanisms on soft materials.

Soft, stretchable rubber capacitors are possible candidates for energy harvesting  ~\cite{Pelrine2001,Jean-Mistral2008,Graf2010a,Brochu2010}, 
that have already been tested in a realistic ocean setting~\cite{Chiba2007,Chiba2008}. 
They were originally introduced as actuators~\cite{Pelrine2000b,Carpi2008,Brochu2010a,Carpi24122010}, 
capable of high actuation strains of more than 100\% and stresses of more than
\unit[1]{MPa}. With a soft capacitor, mechanical energy can be used to pump
charges from a low electrical potential $U$ to a higher, such that the electrical energy
difference can be harvested~\cite{Pelrine2001}. This is made possible by the
large changes of capacitance under mechanical deformation. Although the method is
simple and proven~\cite{Pelrine2001,Chiba2007,Chiba2008,Jean-Mistral2008,Graf2010a,Brochu2010}, 
it is still not clear to which extent the approach is practically useful, 
which is the concern of this paper. Of the many electro-active polymers, it appears that soft capacitors could have the highest energy densities~\cite{Jean-Mistral2010}.

For the purpose of a broad and realistic investigation, a minimal yet realistic 
model is proposed that takes into account the mechanical and dielectric properties 
of the soft capacitor material, including losses and limiting criteria. 
The model also includes cyclic mechanical driving, and an electrical control mechanism 
consisting of a switchable electrical circuit.
The quality of the energy harvesting is characterized by efficiency and gain measures, 
which were evaluated for simulations on a very large set of varied system and material parameters.


The so far only commercially available soft capacitor material, DEAP
(Danfoss PolyPower A/S), is used here as benchmark~\cite{Benslimane2002,DanfossPolyPower,Benslimane2010}. DEAP consists of a sheet of silicone elastomer with smart
compliant metal electrodes and is a realistic candidate with adequate properties
for energy harvesting~\cite{Graf2010}.
The high voltage required for operation of soft capacitor actuators has led to research efforts aiming at lowering the voltages by modifying their mechanical or dielectric properties. Preferably, they should have low mechanical
stiffness and high dielectric constant,
$\varepsilon_r$~\cite{Carpi2008a,Molberg2010,Gallone2010,Stoyanov2010,
StoyanovC0SM00715C}. In general, the dielectric constant can be adjusted from 2 to more
than 1000, however, at higher values such strategies usually cause excessive
conductivity and premature electrical breakdown.

\section{The Model}

The model describes the dynamics of the deformations of and the voltage across the soft capacitor. It includes mechanical and dielectric properties. Losses appear mainly electrically in the electrodes and in the charging and discharging
circuits. The external force is sinusoidal and linearly coupled to the length of the soft capacitor; 
we regard these as minimum requirements for the description of coupling to near-coast oceanic surface waves. 
Technically, elaborate wave-capacitor coupling mechanisms are
possible, yet here the focus is placed on the general principle. 
Mechanical conversion efficiency and electrical
energy gain factors  are calculated from time-integrated losses, mechanical 
input and electrical output. 

The deformation of the polymer film is described in terms of the deformation
ratios $\lambda_i = L_i / L_i^\prime$ (with $i=1,2,3$ for the
$x,y,z$-axes), where $L_i$ and $L_i^\prime$ are the final and the initial
dimensions, respectively. The electrical field is applied in
$z$-direction and the mechanical driving force $f_1$ acts in
$x$-direction. For elastomers, which are amorphous and non-crystalline, wolume conservation can be assumed, $\lambda_1 \lambda_2
\lambda_3 = 1$. Also, the pure shear condition is assumed, $\lambda_2=\text{constant}$.

The time-dependent mechanical response is chosen as a
Kelvin-like fluid~\cite{Brinson-Polymer-Book},
\begin{equation}
T_{ii} = p + \frac{f_i}{A_i} - \gamma \dot{\lambda}_i
\,\text{,} \label{eq:viscous_stress}
\end{equation}
which balances the internal material stress $T_{ii}$, the
hydrostatic pressure $p$, the external forces $f_i$, and the viscous
damping characterized by $\gamma$. Eliminating the pressure
from both equations, utilizing the volume constraint ($\lambda_3 =
\lambda_1^{-1}$), and splitting the material stress into a mechanical
and an electrical part~\cite{Kofod2005} yields
\begin{equation}
- \gamma \dot{\lambda}_1 = 
\left[ \frac{\lambda_1^2 }{\lambda_1^2 + 1} \right]
\left( 
\sigma_{\rm Mech} + \sigma_{\rm Elec} -
\frac{\lambda_1 f_1}{L_2^\prime L_3^\prime }
\right)
\,\text{.}
\label{eq:ode_lambda_ps}
\end{equation}
For 
$\sigma_{\rm Mech}$, 
the Yeoh model~\cite{Yeoh-93} is chosen, with parameters $C_{10}$, $C_{20}$, $C_{30}$:
\begin{equation}
\sigma_{\rm Mech} = 2 \left( \lambda_1^2 - \lambda_1^{-2} \right)
\left( C_{10} + 2 C_{20} \Lambda + 3 C_{30} \Lambda^2 \right)\; ,
\end{equation}
where $\Lambda = \lambda_1^2 + \lambda_1^{-2}-2$. The Young's modulus
of the Yeoh material under pure shear conditions is $Y=8 C_{10}$.

The electrical part of the internal stress $\sigma_{\rm Elec}$, also known as  
Maxwell stress is $\sigma_{\rm Elec}=
-\varepsilon_r \varepsilon_0 E^2$, where $E = U/L_3$ is the electrical
field between the electrodes~\cite{Pelrine2000b,Kofod2005,Suo2008}. It describes
the stress due to the voltage difference $U$ between the capacitor
electrodes. The permittivity of an elastomer does not change at the relatively low levels of electrical field encountered here, and it also does not change when it is deformed, as has been repeatedly verified experimentally by researchers \cite{Lillo2011,Kofod2005}.

The dynamics of the voltage between the electrodes can be derived from $Q = C U$, where $Q$ is its charge and $C$ the capacitance. Building the time derivative and rearranging this equation yields
\begin{equation}
\dot{U} = -\frac{\sigma_{\rm DC} }{ \varepsilon_0 \varepsilon_r} U +
\frac{I_{\rm C}}{C' \lambda_1^2} - 2U
\frac{\dot{\lambda}_1}{\lambda_1}
\,\text{.}\label{eq:ode_voltage_final}
\end{equation}
The first term corresponds to leakage current due to the finite
polymer conductivity $\sigma_{\rm DC}$, the second 
accounts for the electrical control circuit, and the
third for mechanical deformation. $C^\prime = \varepsilon_r
\varepsilon_0 L_1^\prime L_2^\prime / L_3\prime $ is the capacitance of the
unperturbed capacitor, and $C=C^\prime \lambda_1^2$ that of the deformed.

\begin{figure}
  \begin{center}
    \includegraphics[draft=false]{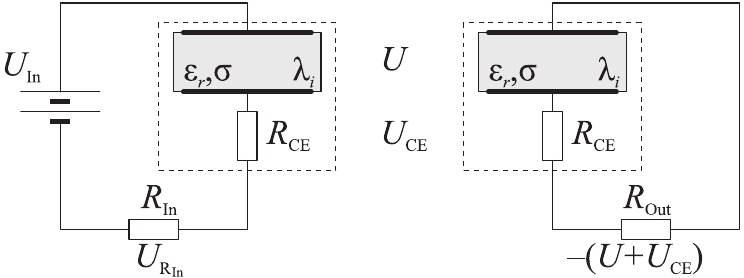}
  \end{center}
  \caption{\label{fig:circuits}The electrical circuits. Left: charging
process, right: discharging process.}
\end{figure}

Eqs.~\eqref{eq:ode_lambda_ps} and \eqref{eq:ode_voltage_final} define a system of two coupled ordinary differential equations (ODEs) for the stretch $\lambda$ and the voltage $U$. The only unknowns are the external driving force $f_1$ and the electrical current from the charging and discharging circuit. The force is chosen by a simple periodic driving $f_1 = f_{0} + f_{\rm A} \sin ( \omega t + \varphi )$, where the actual geometry of the pick-up mechanism can help in adjusting bias force $f_0$ and amplitude $f_{\rm A}$. The angular frequency $\omega = 2\pi/t_{\rm C}$ is determined by the period of the oscillating force, $t_{\rm C}$.

A very simple charge and discharge control mechanism is chosen here. It is described by three points in time. The voltage is ramped on the soft capacitor at $t_0=0$, stopped at $t_1$, while discharging starts at $t_2$. The discharging process ends at $t_C$. A phase difference, $\varphi$, adjusts the onset of the control relative to the driving force.

The capacitor is charged during $(t_0,t_1)$ in an electrical circuit shown in Fig.~\ref{fig:circuits}. 
Kirchhoff's laws state that $U_{\rm In} + U + U_{{\rm CE}} + U_{\rm R_{In}} = 0$ and that $I_{\rm C}=\text{const}$. $U_{\rm In}$ is the driving voltage of the external voltage supply, $U_{{\rm CE}}$, and  $U_{\rm R_{In}}$ are the voltage drops over the capacitor electrodes and the charging resistor, with resistance $R_{\rm CE}$ and $R_{\rm In}$. This yields the current
\begin{equation}
  I_{\rm C} = - \frac{U_{\rm In} + U }{R_{\rm CE} + R_{\rm In} }
  \,\text{.}
  \label{eq:circuit_loading_current}
\end{equation} 
which follows the voltage ramp  $U_{\rm In}(t) = \frac{t}{t_1}
U_{\rm In,max}$, 
where $U_{\rm In,max}$ is the maximal charging voltage.

In the interval $(t_1,t_2)$ the circuit is detached. Mechanical deformation and variation in voltage takes place due to the external variation in force. The response is influenced by viscoelastic damping, while current leakage leads to partial charge loss.

During $(t_2,t_C)$ 
the discharging electrical circuit is attached, cf. Fig.~\ref{fig:circuits}. 
Now, no external voltage is applied and the discharging resistor $R_{\rm Out}$ is used:
\begin{equation}
  I_{\rm C} = - \frac{U} {R_{\rm CE} + R_{\rm Out} }
  \,\text{.}
\end{equation}
After discharging the cycle starts over.

In the ODEs \eref{eq:ode_lambda_ps} and \eref{eq:ode_voltage_final}
four relaxation processes with different time
scales are given: one mechanical, $\tau_{M}$, two electrical,
$\tau_{C}$ and $\tau_{D}$, and $\tau_{\rm PC}$ describing the
loss of charge through the polymer material. In addition, there is
the period of the driving, $t_C$. The mechanical relaxation of the polymer to an
equilibrium
state $\lambda_1^\star$ is described by $\tau_{M}^{-1} = \de
F(\lambda_1) / \de \lambda_1 |_{\lambda_1^\star}$. $F(\lambda)$ is
given by \eqref{eq:ode_lambda_ps} via $\dot{\lambda}_1 =
F(\lambda_1)$. For the Yeoh model this
time scale is approximately $\tau_{M}=\unit[0.55]{s}$. The electrical relaxation
time scales during
charging and discharging are $\tau_{C} = (R_{\rm CE} + R_{\rm In} ) C$
and $\tau_{D} = ( R_{\rm CE} + R_{\rm Out} ) C$. They are typically in the order
of $\unit[10^{-5}]{s}$. 
The largest time scale, which we term the Maxwell time, describes the loss of charge through
the polymer material
$\tau_{\rm PC} = \frac{\varepsilon_0 \varepsilon_r }{\sigma_{\rm DC}}$.

Material failure sets limits which are monitored during simulations and which have been taken into account: i) $\lambda_1 < 5$ avoids rupture, ii) $T_{11} > 0$ ensures a taut material, iii) $E =
\frac{U}{L_3} < \unit[20]{\frac{V}{m}}$ avoids intrinsic electrical breakdown, and iv)
$\frac{\varepsilon_0 \varepsilon_r V^2}{(L_3^\prime)^2}
\lambda_1^4 < H(\lambda_1)$ avoids the the electromechanical instability, analogous to pull-in failure~\cite{Zhao2007,H-Formula}.

\begin{table}
 \caption{\label{tab:parameters}Default parameters for energy harvesting
simulations.}
 \begin{center}
  \begin{tabular}{l@{\hspace{6ex}}r}
   \hline
   \multicolumn{2}{l}{\bf Material and capacitor parameters } \\
   $L_1^\prime$, $L_2^\prime$, $L_3^\prime$ & $\unit[0.1]{m}$,
$\unit[0.2]{m}$, $\unit[5\cdot10^{-5}]{m}$  \\
   $C_{10}$, $C_{20}$, $C_{30}$ & $\unit[139840]{Pa}$,
$\unit[-6570]{Pa}$, $\unit[1057]{Pa}$ \\
   $\varepsilon_r$ & $\unit[3.1]{}$ \\
   $\sigma_{\rm DC}$ & $\unit[10^{-13}]{\frac{A}{Vm}}$ \\
   $R_{\rm CE}$ & $\unit[1000]{\Omega}$ \\
   $\gamma$ & $\unit[10^6]{s/Pa}$ \\
   \hline 
   \multicolumn{2}{l}{\bf Driving parameters } \\
   \multicolumn{2}{l}{$f_{A}=\unit[1]{N}$, $f_{0}=\unit[2]{N}$,
$t_C=\unit[10]{s}$} \\
   \hline 
   \multicolumn{2}{l}{\bf Circuit parameters } \\
   $U_{\rm In,Max}$ & $\unit[651]{V}$ \\
   $R_{\rm In}$, $R_{\rm Out}$ & $\unit[10^6]{\Omega}$,
$\unit[3.83\cdot10^7]{\Omega}$ \\
   $t_1$, $t_2$, $t_C=2\pi / \omega$ & \\
   \hline
   \multicolumn{2}{l}{\bf Limiting parameters} \\
   $\lambda_{\rm Max}$ & $5$ \\
   $E_{\rm Max}$ & $\unit[20]{V m^{-1}}$ \\
   \hline
  \end{tabular}
 \end{center}
\end{table}

The mechanical work is generally defined as $W = L_1^\prime \int
f(\lambda_1,t) \dot{\lambda}_1 \de t$, where $f(\lambda,t)$ is an
arbitrary force acting on the film. Using this equation and
writing all terms in Eq.~\eqref{eq:ode_lambda_ps} as forces
allows the calculation of the work, for example
$W_{\rm Mech} = L_1^\prime \int A_1 \sigma_{\rm Mech} \dot{\lambda}_1
\de t$.  Energy is conserved over one cycle, hence
$W_{\rm Visco} + W_{\rm Driving} = W_{\rm Mech} + W_{\rm Maxwell}$. The electrical work is similarly $W = \int U I \de t$;
again, several contributions should be evaluated. Energy conservation
leads to $W_{\rm Capacitor} + W_{\rm U,In} + W_{\rm R,In} +
W_{\rm R,CE} = 0$ for the charging, and to $W_{\rm Capacitor} +
W_{\rm R,Out} + W_{\rm R,CE} = 0$ for the discharging. It is emphasized that the definition of the voltage allows for negative or positive values for the work done.

The quality of the harvesting process is evaluated by the harvesting efficiency $\eta$ and the gain $G$. The efficiency $\eta$ is the ratio of electrical output energy to the total
mechanical and electrical input energies (note that $W_{\rm In,Elec} < 0$ by
convention):
\begin{equation}
\eta = \frac{ W_{\rm R,Out} }{ W_{\rm Driving} - W_{\rm U,In}}
\,\text{.}\label{eq:harvest_efficiency}
\end{equation}
The gain $G$ is the net relative electrical energy gained compared to electrical energy invested:
\begin{equation}
 G = -\frac{ W_{\rm R,Out} }{W_{\rm U,In}} - 1
 \,\text{.}\label{eq:gain}
\end{equation}
$G$ is positive (negative) if electrical energy is gained (lost).
Losses occur due viscoelasticity, leakage currents and 
finite resistance of capacitor electrodes and charging circuits.
Both measures are essential to feasibility evaluation of energy conversion in practice.

Eqs.~\eqref{eq:ode_lambda_ps} and
\eqref{eq:ode_voltage_final} can only be solved numerically.
The ratio of smallest to largest time scales is on the order of
$10^6$, making the system stiff and solvable by methods such as the Radau, Rosenbrock,
or Runge-Kutta solvers~\cite{Hairer-Solving-ODE-II,Numerical-Recipes}. Here, the Runge-Kutta 4th order method is used, with an integration time step smaller than the smallest time scale of the system.

\section{Simulation, optimization and results}

\begin{figure}
 {
  \centering
\includegraphics[draft=false,width=0.35\textwidth]{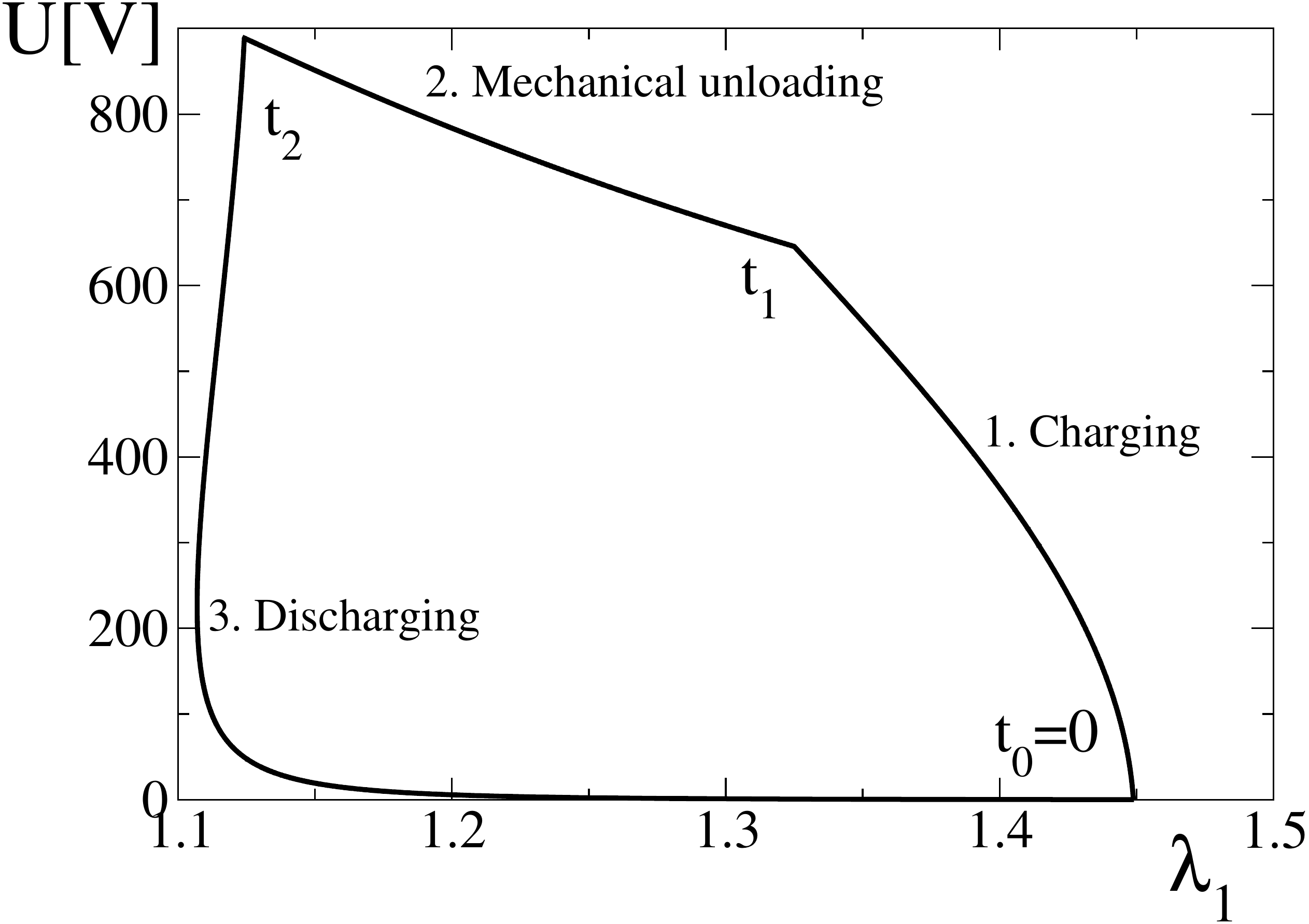}
  \caption{\label{fig:cycle_example}
  An example of a cycle with the default values listed in
Table~\ref{tab:parameters}. The phase of the driving force is
$\varphi=1.75929$ and the switching times are $t_1 = \unit[1.8]{s}$
and $t_2 = \unit[4.32]{s}$.}
 }
\end{figure}

Now follows a presentation of results obtained for the variation of the parameters of 
the harvesting cycle, the electrical circuit, and the materials. Some such parameters could be varied
easily in an operating wave power plant, while others would be fixed by technical and design choices.

The voltage-stretch curve for a simulation using the parameters listed in
Table~\ref{tab:parameters} are shown in Fig.~\ref{fig:cycle_example}. This particular example corresponds to fully optimized parameters for the DEAP material, which were found through procedures described in the sections below.
The mechanical and dielectric parameters are taken from measurements
performed on the DEAP soft capacitor material, and are
consistent with the corresponding data
sheets~\cite{DanfossPolyPower}. The phase of the driving is
$\varphi=1.75929$, the switching times are $t_1 = \unit[1.8]{s}$
and $t_2 = \unit[4.32]{s}$. 
The efficiency in this particular example is $\eta = 0.42$ while the gain is
$G=0.47$.
The total harvested energy is $W_{\rm
R,Out}=\unit[5.5]{mJ}$. Compared to the mass of the material (the
volume is $L_1^\prime L_2^\prime L_3^\prime = \unit[10^{-6}]{m^3}$,
while the density is $\unit[1100]{kg\,m^{-3}}$), the specific
harvested energy for one cycle becomes $\unit[5]{J kg^{-1}}$. 
Of course, this is small
compared to energy density of $\unit[1500]{J kg^{-1}}$ (from~\cite{Pelrine2001}), 
which applies to the special case of prestretched polyacrylate glue. 
However, the result obtained here is more realistic; it is found by considering a full cycle, maximizing $\eta$ and  with a limitation of the electrical field to a reasonable level of $\unit[20]{Vm^{-1}}$.

\begin{figure}
{
\centering
\includegraphics[draft=false,width=0.45\textwidth]{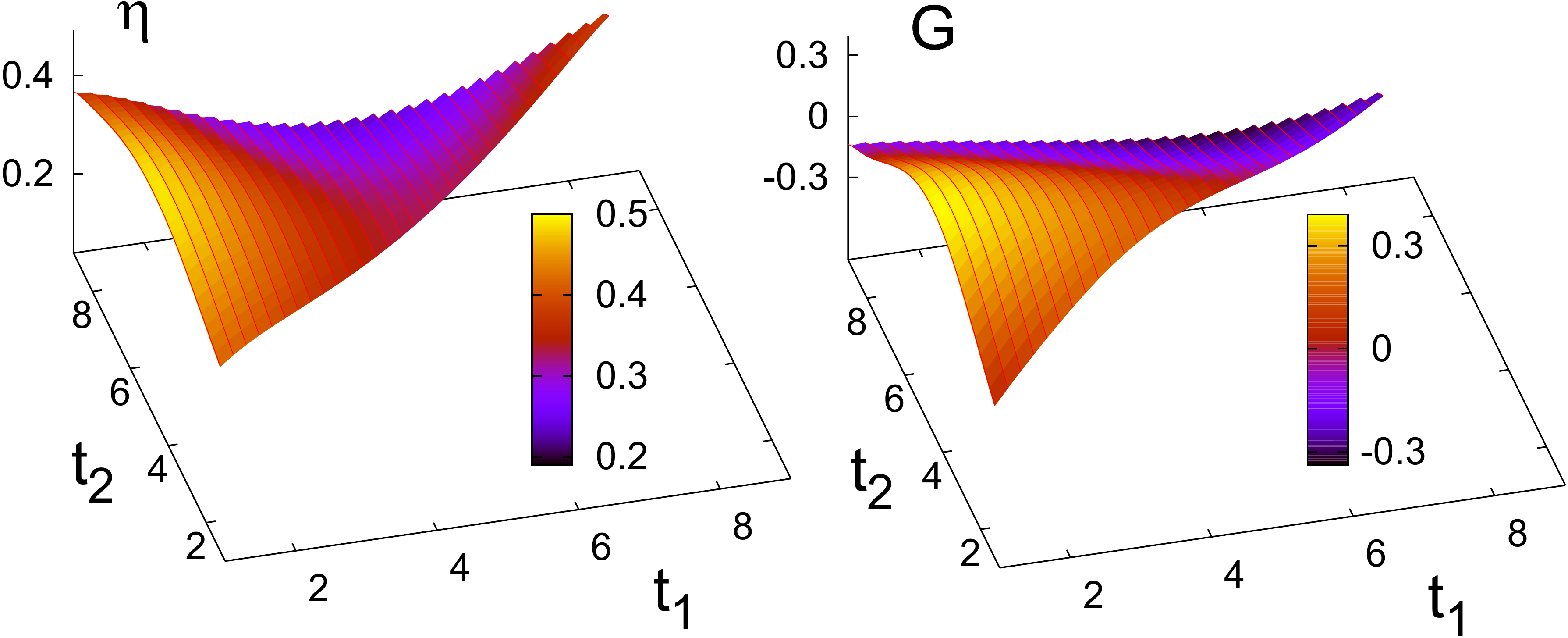}
\caption{
\label{fig:efficiency_times}
The harvesting efficiency $\eta$ and the gain $G$ in dependence of the
switching times $t_1$ and $t_2$. The surfaces are triangular because
$t_1 < t_2$. The phase is fixed $\varphi=1.75929$.}
}
\end{figure}

\subsection{Cycle parameter optimization}

The harvesting cycle allows a variation of the switching times $t_1$, $t_2$
and the phase $\varphi$. Their impact on the optimal efficiency $\eta$ and gain
$G$ is shown in Fig.~\ref{fig:efficiency_times} for the DEAP material 
(see Tab.~\ref{tab:parameters}); the phase is 
fixed in this plot to $\varphi=1.76$. As expected, efficiency and 
gain depend strongly on the cycle parameters. Note
also that regions of maximum gain and maximum efficiency do not
overlap perfectly. Hence, an optimization strategy for efficiency
could result in zero energy gain and vice versa.

In the following, both efficiency {\it and} gain are optimized in
dependence of the circuit and material parameters. For each parameter set, 
$t_1$, $t_2$ and $\varphi$ are varied to maximize the efficiency and the gain, 
either by a Monte Carlo method using random sampling or the 
simplex method~\cite{Numerical-Recipes}.

\begin{figure}[!b]
 \begin{center}
  \includegraphics[draft=false,width=0.45\textwidth]{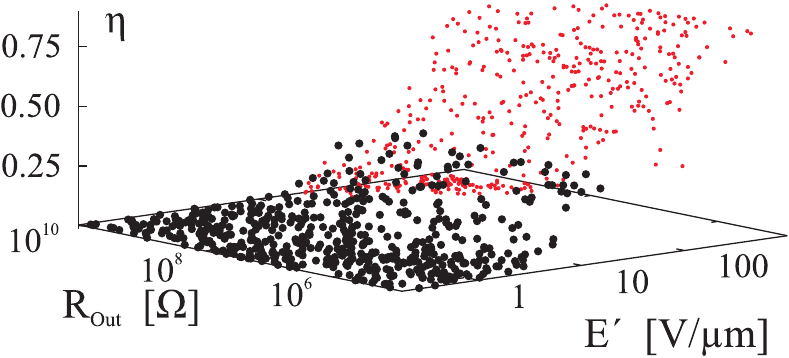}
  \caption{The efficiency in dependence of $E^\prime=U_{\rm In,Max} /
L_3^\prime$ and $R_{\rm out}$. The values of $t_1$, $t_2$ and
$\varphi$ have been optimized with help of the simplex method. Cycles
violating the breakdown criteria have are plotted in red.
  \label{fig:optimize_uin_rout}
  }
 \end{center}
\end{figure}

\subsection{Electrical circuit optimization}

In a power plant setting, the parameters $U_{\rm In,Max}$ and $R_{\rm Out}$ 
are adjustable.
So is $R_{\rm In}$, though this would typically increase loss, 
and thus is not studied further here. 
Fig.~\ref{fig:optimize_uin_rout} illustrates the result of this optimization. It was
obtained by first choosing random
values for $E^\prime=U_{\rm In,Max}/L_3^\prime$ and $R_{\rm Out}$ in
the ranges shown, then optimizing the efficiency by adjusting $t_1$,
$t_2$, $\varphi$ via the simplex method. If the breakdown criteria are
violated, the particular data point is plotted in red. 
The material properties are again taken from Tab.~\ref{tab:parameters}. 
Clearly, the efficiency increases if the driving voltage increases. 
It is bounded by the breakdown criteria, 
and it is relatively unaffected by the output
resistance in the investigated range. The maximum observed harvesting
efficiency was about 0.42.

\begin{figure}
 \begin{center}
  \includegraphics[draft=false,width=0.45\textwidth]{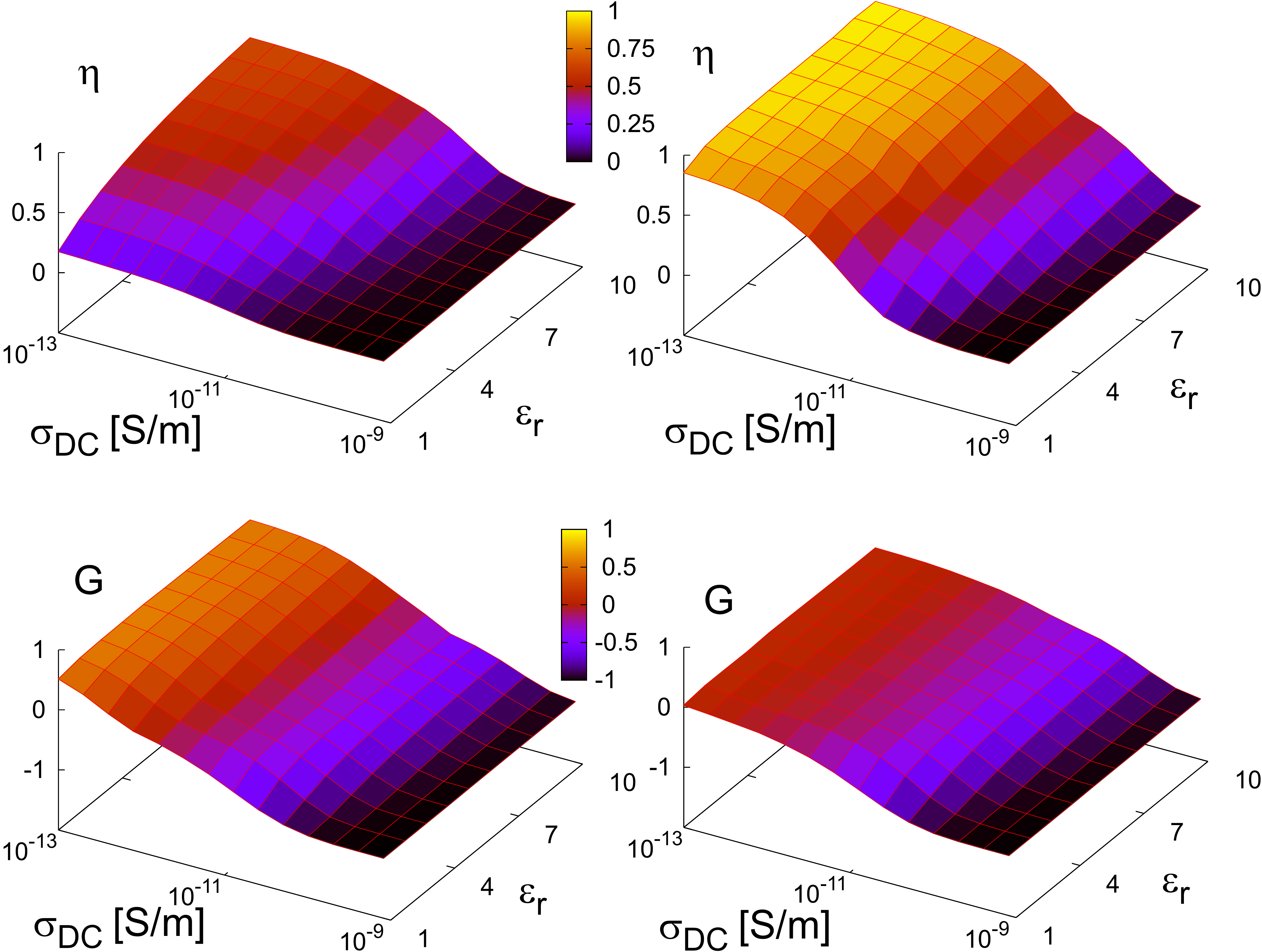}
  \caption{\label{fig:efficiency_eps_sigma} Evaluation of the effect
of material properties. The upper and lower rows show optimized
efficiency $\eta$ and gain $G$, respectively, in dependency of
relative permittivity $\varepsilon_r$ and polymer conductivity
$\sigma_{\rm DC}$. The graphs in the left column was for polymer
modulus $Y=\unit[1.2]{MPa}$, in the right for $Y=\unit[5]{MPa}$. }
 \end{center}
\end{figure}

\subsection{Dielectric material parameter optimization}

Fig.~\ref{fig:efficiency_eps_sigma} shows the numerically optimized
efficiency for varying polymer conductivity $\sigma_{\rm DC}$ and
permittivity $\varepsilon_{\rm rel}$ for two different values of the
stiffness of the polymer. The stiffness is varied by scaling of
$C_{10}$, $C_{20}$, and $C_{30}$. For each individual parameter set
($\varepsilon_r$, $\sigma_{\rm DC}$) 1000 randomized sets of values of
$U_{\rm In}$, $R_{\rm Out}$, $t_1$, $t_2$ and $\varphi$ were
generated; other parameter are held fixed, cf.
Table~\ref{tab:parameters}. Out of these data sets, the one with 
the highest value of $\eta$ (or $G$) not violating the limit criteria was
determined and plotted (therefore,
the observed $\eta$ and $G$ do not correspond to identical $U_{\rm
In}$, $R_{\rm Out}$, $t_1$, $t_2$ and $\varphi$ values).

The results in Fig.~\ref{fig:efficiency_eps_sigma} confirm
intuition: First, $\eta$
generally decreases with $\sigma_{\rm DC}$ and increases with
$\varepsilon_r$. As has been suspected, but not clearly demonstrated
before, it is found here that the stiffer material can achieve a
higher efficiency, for identical $\sigma_{\rm DC}$ and
$\varepsilon_r$. Gain $G$ and efficiency $\eta$ behave similarly,
but  now the stiffer material has the lower response, because they undergo
smaller strains and thus smaller relative change in the capacitance.
The drop in gain appears less pronounced than the increase in
efficiency, which indicates that stiffer
elastomers are advantageous for energy harvesting.

\begin{figure}
{
 \centering
 \includegraphics[draft=false,width=0.45\textwidth]{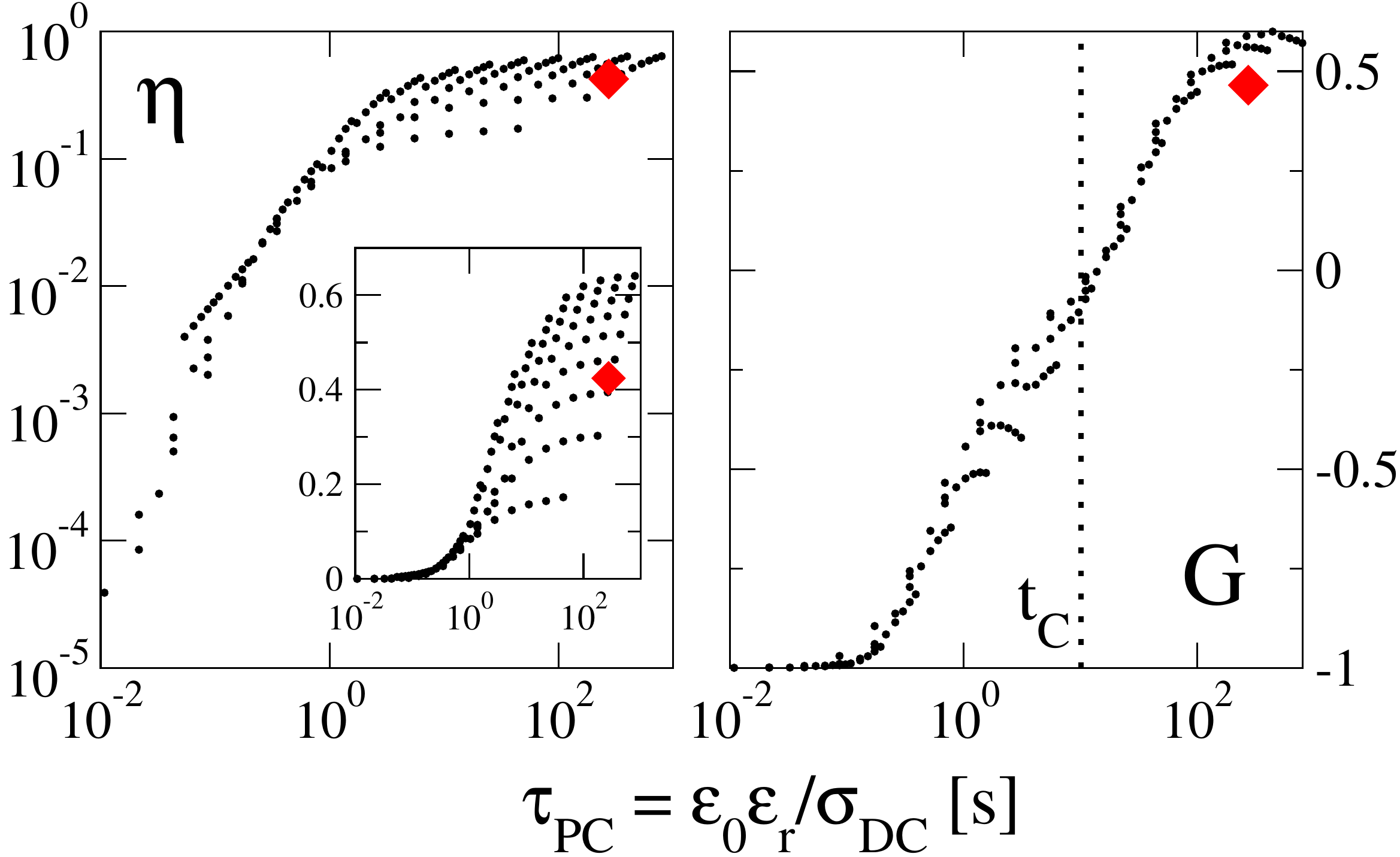}
 \caption{\label{fig:efficiency_maxwell_time} 
The efficiency $\eta$ and the gain $G$ in dependence of the Maxwell time
$\tau_{\rm PC}$. The data are identical to
Fig.~\ref{fig:efficiency_eps_sigma} with $Y = \unit[1.2]{MPa}$. The inset shows a semilogarithmic plot for $\eta$. The big red diamonds show the results for the DEAP material.}
}
\end{figure}

Fig.~\ref{fig:efficiency_eps_sigma} shows also that both $\eta$ and
$G$ depend roughly on $\varepsilon_r \times \sigma_{\rm DC}^{-1} $. We
note that this is nearly the expression for the time constant of the
charge decay in the capacitor, $\tau_{\rm PC} = \varepsilon_r
\varepsilon_0 /\sigma_{\rm DC}$,  also known as the Maxwell
relaxation time. To elucidate this observation, $\eta$ and $G$ are plotted against
$\tau_{\rm PC}$, see Fig.~\ref{fig:efficiency_maxwell_time}. All efficiency values appear to fall below a particular threshold; below $\tau_{\rm PC} = t_C$, the efficiency varies with $\tau_{\rm PC}$, above it is nearly constant and are distributed between $0.2$ and $0.6$.

Also gain is seen to depend directly upon $\tau_{\rm PC}$. It collapses on a straight line in the semilogarithmic plot. Interestingly, it is only positive for $\tau_{\rm PC} > t_C$, clearly showing that the Maxwell relaxation time must be shorter than the period of the mechanical driving. Comparing to the efficiency plot, high gain does not necessarily lead to high efficiency as was seen previously, see Fig.~\ref{fig:efficiency_times}. In summary, the highest efficiencies and gains are slightly above 0.6 and encouragingly, the DEAP material shows both good efficiency {\em and} gain.

These plots clearly show the importance of the proper choice of the time scales of the loss mechanisms, and that they must be chosen in accordance with the relevant driving time scale. Furthermore, $\tau_{\rm PC}$ can be identified as the dominant material parameter for energy harvesting.

\begin{figure}
{
 \centering
\includegraphics[draft=false,width=0.48\textwidth]{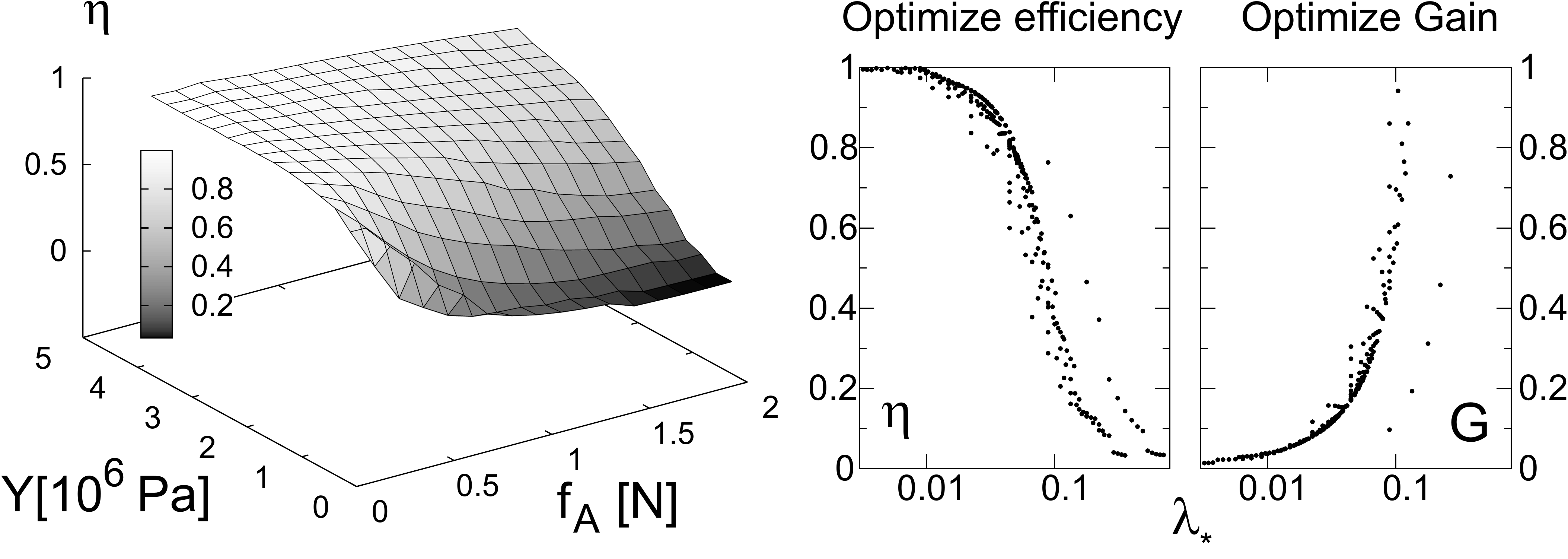}		
 \caption{\label{fig:efficiency_f_c10}\label{fig:lambdac_efficiency_gain}Left
panel: the efficiency $\eta$ in
dependence of the amplitude of the driving force $f_{A}$ and the
stiffness $Y=8C_{10}$ of the material. Right panel: The efficiency and
the gain in dependency of the $\lambda_* = \sigma_{\rm Amp}/Y$. The
electrical material parameter are taken from~\ref{tab:parameters}.}
}
\end{figure}

\subsection{Mechanical material parameter optimization}

The impact of the mechanical properties of the elastomer was evaluated
by varying the stress-strain properties, simply by multiplication of
$C_{10}$, $C_{20}$, and $C_{30}$ with the same factor. Then, the
strength of the driving force was varied and either the efficiency
$\eta$ or the gain $G$ was optimized, as described above. As is clearly seen from Fig.~\ref{fig:efficiency_f_c10} (left), the efficiency increases with higher stiffness,
while it decreases for higher driving forces.

This data is efficiently analyzed when introducing a new variable,
the virtual stretch $\lambda_* = \sigma_{\rm Amp} / Y = f_A / ( L_2^\prime
L_3^\prime / Y)$, which is a
relative measure of the level of stretch during the cycle (see
Fig.~\ref{fig:lambdac_efficiency_gain}). Remarkably, this plot shows
that small stretches (originating from high stiffnesses or small
driving forces) result in high efficiency. The gain behaves opposite,
therefore energy harvesting can only be practicable in a narrow range of $\lambda_*$.

\section{Conclusion}

In this article, energy harvesting using electro active polymers has been studied by means of numerical simulations and analytical considerations. In detail, a realistic model describing the dynamics of the polymer as a relaxation process has been presented, as well as electrical control circuits and various limiting criteria. This system is driven by a periodic force, which compares well to an energy harvesting device using ocean waves or similar (fluid) mechanical forcing, as in generators in shoes or energy producing patches. The model is generic in that all its units describe essential and necessary parts of any future realization of an energy harvesting power plant. A direct coupling between the driving force and the soft capacitor was chosen, which applies well if complicated and expensive mechanical setups have to be avoided.

The study focused on qualitative and quantitative measures (efficiency and gain) for energy harvesting and on the corresponding parameter optimization, which is important for the design of potential power plants. The study showed that optimizing charging and discharging times as well as the electrical loading parameters are crucial to obtain high efficiencies.  Furthermore, it could be shown that the Maxwell time $\tau_{\rm PC}$ and the virtual stretch $\lambda_*$ are powerful tools in the analysis and should be tuned rationally to obtain optimal energy output. Among others, this leads to the strong conclusion that the Maxwell time must by larger than the period of the mechanical driving.

The realistic simulation of the DEAP silicone material shows a specific harvesting energy density of $\unit[5]{J kg^{-1}}$ for a wave period of \unit[10]{s}. Assuming a realistic raw material price for silicone of $\unit[10]{Euro/kg}$, this study indicates that 
a full return of materials investment is possible within 10 years. 
The investment will of course be compounded by other costs, for electronics, installations and maintenance.
This time span is a very conservative estimate. It can be reduced if electrical fields larger than $\unit[20]{V \mu m^{-1}}$ are used, which could be possible with stiffer materials~\cite{Kollosche2009}. Further improvements should be possible with a more elaborate charging and discharging scheme that exploits the full variable range within the limit criteria~\cite{Graf2010,Jean-Mistral2010}. As such, this study shows that the wave energy harvesting technology based on soft capacitors has great potential for practical usability.

\section{Appendix: Mechanical model}

The time dependent mechanical response of the polymer is given by Eq.~\eqref{eq:viscous_stress} which describes a Kelvin-like fluid. It assumes that shear stresses are not present and that the material will always relax back to the equilibrium. $\gamma$ is strongly related to the mechanical relaxation time. In fact, the mechanical relaxation $\tau_M$ is given by $\tau_M^{-1}=\de F(\lambda_1) / \de \lambda_1 |_{\lambda_1^\star} \approx Y /  2 \gamma$, where $F(\lambda_1)$ is the RHS of \eqref{eq:ode_lambda_ps}. A value $\gamma=\unit[10^6]{sPa^{-1}}$ was chosen resulting in a relaxation time of $\tau_{M} \approx \unit[1.8]{s}$ if external forces and charges are absent.

To derive the final constitutive differential equation, the equations for $T_{11}$ and $T_{33}$ are considered 
\begin{subequations}
\begin{eqnarray*}
T_{11} & = & \sigma_{11}^M + \sigma_{11}^E - p = \frac{f_1}{A_1} -
\gamma \dot{\lambda}_1
\\
T_{33} & = & \sigma_{33}^M + \sigma_{33}^E - p = - \gamma \dot{\lambda}_3
\,\, \text{,}
\end{eqnarray*}
\end{subequations}
and the static pressure $p$ is eliminated to obtain
\begin{equation}
\left( \sigma_{11}^M - \sigma_{33}^M \right) +
\left( \sigma_{11}^E - \sigma_{33}^E \right) = \frac{f_1}{A_1} -
\gamma \left( \dot{\lambda}_1 - \dot{\lambda}_3 \right)
\,\,\text{.}
\label{eq:stress_balance}
\end{equation}
The force is applied in $x$-direction. Furthermore, $T_{22}$ does not enter the game since pure-shear boundary conditions are assumed.

The term $\sigma_{11}^M - \sigma_{33}^M = \sigma_{\rm Mech}$ is the stress strain model and depends in general on the geometry of the polymer. The Yeoh model was chosen which agrees well with experimental results on the Danfoss material. It can be derived from $\sigma_{ii} = \lambda_i \partial W / \partial \lambda_i$. $W$ is the energy function which takes the form $W=C_{10} ( I_1 - 3 ) + C_{20} ( I_1 - 3 ) ^2 + C_{30} ( I_1 - 3 )^3$ for the Yeoh model with $I_1 = \lambda_1^2 + \lambda_2^2 + \lambda_3^2$. Inserting the pure-shear assumptions and summarizing the terms for the $x$ and the $z$ directory lead finally to 
\begin{equation}
\sigma_{\rm Mech} = 2 \left( \lambda_1^2 - \lambda_1^{-2} \right)
\left( C_{10} + 2 C_{20} \Lambda + 3 C_{30} \Lambda^2 \right)
\end{equation}
with $\Lambda = \lambda_1^2 + \lambda_1^{-2}-2$. Usually the stiffness of materials is quantified by the Youngs modulus $E$, which is defined as $Y = \frac{\partial \sigma}{\partial \lambda}\big|_{\lambda=1}$. For the Yoeh model under pure shear conditions this results in $Y=8 C_{10}$.

The second term on the LHS of \eref{eq:stress_balance} is the Maxwell stress \cite{Kofod2005,Suo2008}
\begin{equation}
\sigma_{\rm Elec} = \sigma_{11}^E - \sigma_{33}^E = - \varepsilon_r \varepsilon_0 E_{3}^2
= -\varepsilon_r \varepsilon_0 \frac{U^2}{L_3^{\prime2}} \lambda_1^2
\,\,\text{,}
\end{equation}
where $E_{3}$ is the electrical field and $U$ the voltage between the electrodes.


\begin{thebibliography}{10}

\bibitem{Cruz-08}
Joao Cruz, editor.
\newblock {\em Ocean Wave Energy}.
\newblock Springer, New York, 2008.

\bibitem{Dalton2010}
G.~Dalton and B.~P. {{\'O} Gallach{\'o}ir}.
\newblock Building a wave energy policy focusing on innovation, manufacturing
  and deployment.
\newblock {\em Renew. Sust. Energ. Rev.}, 14(8):2339 -- 2358, 2010.

\bibitem{Pelrine2001}
Ron Pelrine, Roy~D. Kornbluh, Joseph Eckerle, Philip Jeuck, Seajin Oh, Qibing
  Pei, and Scott Stanford.
\newblock Dielectric elastomers: generator mode fundamentals and applications.
\newblock In Yoseph Bar-Cohen, editor, {\em Smart Structures and Materials
  2001: Electroactive Polymer Actuators and Devices}, volume 4329, pages
  148--156. SPIE, 2001.

\bibitem{Jean-Mistral2008}
Claire Jean-Mistral, Skandar Basrour, and Jean-Jacques Chaillout.
\newblock Dielectric polymer: scavenging energy from human motion.
\newblock volume 6927, page 692716. SPIE, 2008.

\bibitem{Graf2010a}
C.~Graf, M.~Aust, J.~Maas, and D.~Schapeler.
\newblock Simulation model for electro active polymer generators.
\newblock In {\em Proc. ACTUATOR, Bremen}, page P73, 2010.

\bibitem{Brochu2010}
Paul Brochu, Huafeng Li, Xiaofan Niu, and Qibing Pei.
\newblock Factors influencing the performance of dielectric elastomer energy
  harvesters.
\newblock volume 7642, page 76422J. SPIE, 2010.

\bibitem{Chiba2007}
Seiki Chiba, Mikio Waki, Roy Kornbluh, and Ron Pelrine.
\newblock Extending applications of dielectric elastomer artificial muscle.
\newblock In Yoseph Bar-Cohen, editor, {\em Proc. SPIE-EAPAD}, volume 6524,
  page 652424, 2007.

\bibitem{Chiba2008}
Seiki Chiba, Mikio Waki, Roy Kornbluh, and Ron Pelrine.
\newblock Innovative power generators for energy harvesting using electroactive
  polymer artificial muscles.
\newblock In Yoseph Bar-Cohen, editor, {\em Proc. SPIE-EAPAD}, volume 6927,
  page 692715, 2008.

\bibitem{Pelrine2000b}
Ron Pelrine, Roy Kornbluh, Qibing Pei, and Jose Joseph.
\newblock High-speed electrically actuated elastomers with strain greater than
  100\%.
\newblock {\em Science}, 287:836--839, 2000.

\bibitem{Carpi2008}
Federico Carpi, Danilo De~Rossi, Roy Kornbluh, Ronald Pelrine, and Peter
  Sommer-Larsen, editors.
\newblock {\em Dielectric elastomers as electromechanical transducers}.
\newblock Elsevier, Amsterdam, 2008.

\bibitem{Brochu2010a}
Paul Brochu and Qibing Pei.
\newblock Advances in dielectric elastomers for actuators and artificial
  muscles.
\newblock {\em Macromol. Rapid Comm.}, 31(1):10--36, 2010.

\bibitem{Carpi24122010}
Federico Carpi, Siegfried Bauer, and Danilo De~Rossi.
\newblock Stretching dielectric elastomer performance.
\newblock {\em Science}, 330(6012):1759--1761, 2010.

\bibitem{Jean-Mistral2010}
C~Jean-Mistral, S~Basrour, and J-J Chaillout.
\newblock Comparison of electroactive polymers for energy scavenging
  applications.
\newblock {\em Smart Mater. and Struct.}, 19(8):085012, 2010.

\bibitem{Benslimane2002}
Mohammed Benslimane, Peter Gravesen, and Peter Sommer-Larsen.
\newblock Mechanical properties of dielectric elastomer actuators with smart
  metallic compliant electrodes.
\newblock In {\em Proc. SPIE-EAPAD}, volume 4695, pages 150--157, 2002.

\bibitem{DanfossPolyPower}
{Danfoss PolyPower A/S}.
\newblock ``094F0031 Film Kit Engineering Sheet 15-12-2009,'' Technical Data
  Sheet.

\bibitem{Benslimane2010}
Mohamed~Y Benslimane, Hans-Erik Kiil, and Michael~J Tryson.
\newblock Dielectric electro-active polymer push actuators: performance and
  challenges.
\newblock {\em Polymer International}, 59(3):415--421, 2010.

\bibitem{Graf2010}
C.~Graf, M.~Aust, J.~Maas, and D.~Schapeler.
\newblock Simulation model for electro active polymer generators.
\newblock In {\em Proc. 10th IEEE-ICSD, Potsdam}, pages G1--1, 2010.

\bibitem{Carpi2008a}
F.~Carpi, G.~Gallone, F.~Galantini, and D.~{De Rossi}.
\newblock Silicone--poly(hexylthiophene) blends as elastomers with enhanced
  electromechanical transduction properties.
\newblock {\em Adv. Funct. Mater.}, 18(2):235--241, 2008.

\bibitem{Molberg2010}
Martin Molberg, Daniel Crespy, Patrick Rupper, Frank N{\"u}esch, Jan-Anders~E.
  M{\aa}nson, Christiane L{\"o}we, and Dorina~M. Opris.
\newblock High breakdown field dielectric elastomer actuators using
  encapsulated polyaniline as high dielectric constant filler.
\newblock {\em Adv. Funct. Mater.}, 20(19):3280--3291, 2010.

\bibitem{Gallone2010}
Giuseppe Gallone, Fabia Galantini, and Federico Carpi.
\newblock Perspectives for new dielectric elastomers with improved
  electromechanical actuation performance: composites versus blends.
\newblock {\em Polym. Int.}, 59(3):400--406, 2010.

\bibitem{Stoyanov2010}
Hristiyan Stoyanov, Matthias Kollosche, Denis~N. McCarthy, and Guggi Kofod.
\newblock Molecular composites with enhanced energy density for electroactive
  polymers.
\newblock {\em J. Mater. Chem.}, 20:7558--7564, 2010.

\bibitem{StoyanovC0SM00715C}
Hristiyan Stoyanov, Matthias Kollosche, Sebastian Risse, Denis~N. McCarthy, and
  Guggi Kofod.
\newblock Elastic block copolymer nanocomposites with controlled interfacial
  interactions for artificial muscles with direct voltage control.
\newblock {\em Soft Matter}, pages~--, 2010.

\bibitem{Brinson-Polymer-Book}
Hal~F. Brinson and L.~Catherine Brinson.
\newblock {\em Polymer Engineering Science and Viscoelasticity: An
  Introduction}.
\newblock Springer, Berlin, December 2007.

\bibitem{Kofod2005}
Guggi Kofod and Peter Sommer-Larsen.
\newblock Silicone dielectric elastomer actuators: Finite-elasticity model of
  actuation.
\newblock {\em Sens. Actuat. A}, 122(2):273--283, aug 2005.

\bibitem{Yeoh-93}
O.~H. Yeoh.
\newblock Some forms of the strain energy function for rubber.
\newblock {\em Rubber Chem. Technol.}, 66(5):754--771, 1993.

\bibitem{Suo2008}
Zhigang Suo, Xuanhe Zhao, and William~H. Greene.
\newblock A nonlinear field theory of deformable dielectrics.
\newblock {\em J. Mech. Phys. Solids}, 56(2):467 -- 486, 2008.

\bibitem{Lillo2011}
L.~Di Lillo, A.~Schmidt, A.~Bergamini, P.~Ermanni, and E.~Mazza.
\newblock Dielectric and insulating properties of an acrylic dea material at
  high near-dc electric fields.
\newblock volume 7976, page 79763B. SPIE, 2011.

\bibitem{Zhao2007}
Xuanhe Zhao and Zhigang Suo.
\newblock Method to analyze electromechanical stability of dielectric
  elastomers.
\newblock {\em Appl. Phys. Lett.}, 91(6):061921, 2007.

\bibitem{H-Formula}
$H(\lambda_1)= \frac{1}{\lambda_1^4}(42 C_{30}) + \frac{1}{\lambda_1^2}(20
  C_{20} - 120 C_{30}) + 6 C_{10} - 24 C_{20} + 90 C_{30} + \lambda_1^4 (20
  C_{10} - 8 C_{20} - 30 C_{30}) + \lambda_1^6 (12 C_{20} - 72 C_{30}) +
  \lambda_1^8 (30 C_{30})$.

\bibitem{Hairer-Solving-ODE-II}
Ernst Hairer and Gerhard Wanner.
\newblock {\em Solving Ordinary Differential Equations {II:} Stiff and
  {Differential-Algebraic} Problems}.
\newblock Springer, Berlin, February 2010.

\bibitem{Numerical-Recipes}
William~H Press, Saul~A Teukolsky, William~T Vetterling, and Brian~P Flannery.
\newblock {\em Numerical Recipes 3rd Edition: The Art of Scientific Computing}.
\newblock Cambridge University Press, September 2007.

\bibitem{Kollosche2009}
Matthias Kollosche and Guggi Kofod.
\newblock Electrical failure in blends of chemically identical, soft
  thermoplastic elastomers with different elastic stiffness.
\newblock {\em Appl. Phys. Lett.}, 96(7):071904, 2010.

\end{thebibliography}

\end{document}